\UseRawInputEncoding 
\documentclass[11pt]{article}

\usepackage{amssymb}\usepackage[centertags]{amsmath}\usepackage{txfonts}\usepackage{epsfig}\usepackage{bm}
\usepackage{color}\usepackage{graphicx,graphics}\usepackage{multirow}\usepackage{float}\usepackage{ulem}
\usepackage{setspace}\usepackage{slashed}\usepackage{booktabs}\usepackage{hyperref}
\usepackage[text={160mm,225mm},centering]{geometry}
\hypersetup{colorlinks=true, citecolor=blue, linkcolor=red, filecolor=black,urlcolor=blue}
\def\fnote#1#2{\begingroup\def\thefootnote{#1}\footnote{#2}\addtocounter
{footnote}{-1}\endgroup}

\allowdisplaybreaks[4]

\begin{document}


\begin{center}
{\Large {\bf {Charged current semi-inclusive deeply inelastic scattering at the Electron-Ion Collider}}}

\vspace{20pt}

Weihua Yang \fnote{*}{}\\
{\em College of Nuclear Equipment and Nuclear Engineering,\\
 Yan Tai University, Yantai, 264005, China }

\vspace{30pt}

\noindent
{\bf Abstract}
\end{center}

We present a systematic calculation of the charged current semi-inclusive deeply inelastic scattering process at leading order twist-3 level in the parton model. We consider the general form of the negatively charged beam scattering off the polarized target which has spin-1. The calculations are carried out by applying the collinear expansion where multiple gluon scattering is taken into account and gauge links are obtained automatically.
We first present the general form of the differential cross section in terms of the structure functions by kinematic analysis and then present the structure functions in terms of the the gauge invariant parton distribution functions up to twist-3 level. Considering the angle modulations and polarizations of the cross section, we calculate the complete azimuthal asymmetries in the charged current semi-inclusive deeply inelastic scattering process. The charge asymmetries are also considered in this paper with the introduction of the definitions of the $plus$ and $minus$ cross sections.

\vfill

\pagebreak

\section{Introduction}

Factorization theorems \cite{Collins:1989gx} that enable one to apply perturbative calculations to many important processes involving hadrons can separate the calculable hard parts from the nonperturbative soft parts in calculating the cross section in the quantum chromodynamics (QCD). These soft parts are often factorized as parton distribution functions (PDFs) and fragmentation functions which are most easily seen in deeply inelastic scattering (DIS) and electron positron annihilation processes.
The DIS process has proven to be an important tool for exploring the structure of nucleons and even of nuclei. Data from DIS experiments provide important and invaluable information on both the partonic structures and spin structures of nucleons \cite{Donnelly:1975ze,Bloom:1969kc,Ashman:1987hv,Ashman:1989ig}. In this paper, we consider the jet production semi-inclusive DIS (SIDIS) process where both the scattered lepton and the current jet are detected. The calculations are carried out by applying the collinear expansion method which has proven to be a powerful tool in calculating both the DIS process \cite{Ellis:1982wd,Qiu:1990xxa} and jet production SIDIS process \cite{Liang:2006wp}.
When three dimensional, i.e., the transverse momentum dependent (TMD) PDFs are considered in the SIDIS process, the sensitive quantities studied in experiments, are often different azimuthal asymmetries. These asymmetries are measurable quantities which can be used to extract the corresponding TMD PDFs.

In DIS process leptons scattering from nuclei is due to the electromagnetic interaction of the leptons with quarks. Under the one photon approximation condition, the electromagnetic interaction can be described accurately which results in the straightforward interpretation of experiment data. In addition to the electromagnetic interaction, weak interaction in the scattering process provides more physical information, e.g., parity violating asymmetry  \cite{Cahn:1977uu,Anselmino:1993tc,Prescott:1978tm,Prescott:1979dh,Kumar:2013yoa,Aniol:2004hp,Aniol:2005zf, Aniol:2005zg,Armstrong:2005hs,Androic:2009aa, Wang:2013kkc,Wang:2014guo,Anthony:2003ub,Anthony:2005pm,Spayde:1999qg,Ito:2003mr,Maas:2004dh,Maas:2004ta}.
Comparing to the neutral current (NC) SIDIS process which propagate with $\gamma^*$ and/or $Z^0$ boson, data from the charged current (CC) experiments provide more complementary information on the partonic structure of nucleons as they probe combinations of quark flavors different from those accessible in NC ones. Though, it is experimentally difficult to identify the jets from light quarks in jet production SIDIS, the theoretically clean CC channel present stringent tests of SIDIS computations as well as addressing the universality of the PDFs \cite{Aschenauer:2017jsk}.
However, CC experiments can only be studied either in high-energy lepton-nucleon collisions \cite{Aaron:2009aa,Collaboration:2010xc,Aaron:2012qi} or at neutrino scattering experiments \cite{Tzanov:2009zz}.
We now consider the CC SIDIS at the Electron-Ion Collider (EIC)~\cite{Accardi:2012qut} energies in this paper.

The EIC is a high-energy, high-luminosity collider with the capability to accelerate polarized electron and nucleon/ions. It will provide access to kinematic regions where gluons dominate in the nucleon or nuclei and access to the spatial and spin structure of gluons in the proton. Though, the EIC is proposed mainly for understanding the fundamental strong theory of the quark and gluon fields, it opens a new window to measure new physical quantities for weak theory in the scattering experiments.
Neutral current inclusive and semi-inclusive DIS processes have been studied extensively \cite{Ji:1993ey,Anselmino:1994gn,Boer:1999uu,Anselmino:2001ey,deFlorian:2012wk,Moreno:2014kia,Chen:2020ugq}, we pay attention to the calculation of the CC jet production SIDIS process for spin-1 polarized target at the EIC energies in this paper.
The calculations are carried out by applying the collinear expansion formalism which is a powerful tool to calculate higher twist effects systematically by taking into account multiple gluon exchange contributions. By using collinear expansion, on the one hand gauge links will be generated automatically which make the calculation explicitly gauge invariant. On the other hand the formalism takes a very simple factorization form which consists of calculable hard parts and TMD PDFs. This will greatly simplify the systematic calculation of higher twist contributions.

The rest of this paper is organized as follows. In Sec.~\ref{sec:formalism}, we make kinematic analysis for CC SIDIS process and present the general form of the differential cross section in terms of structure functions.
In Sec.~\ref{sec:partonmodel} and \ref{sec:crosssection}, we present detailed calculations of the hadronic tensor and the cross section respectively up to twist-3 level in terms of the gauge invariant TMD PDFs in the parton model.
The results including structure functions, azimuthal asymmetries and charge asymmetries are given in Sec.~\ref{sec:result}.
Finally, a summary is given in Sec.~\ref{sec:summary}.

\section{The general form of the cross section} \label{sec:formalism}

In this section we present the general formalism of the CC SIDIS, including the hadronic tensor decomposed by the basic Lorentz tensors (BLTs) and cross section in terms of the structure functions.
\subsection{The charged current SIDIS process}

To be explicit, we consider the current jet production SIDIS process at EIC energies,
\begin{align}
l(l,\lambda_l) + N(p,S) \rightarrow l^\prime(l^\prime) + q(k^\prime) + X,
\end{align}
where $l$ denotes a electron/positron and $l^\prime$ is the corresponding neutrino, $\lambda_l$ is the helicity of the initial lepton. $N$ can be a nucleon with spin-1/2 or an ion, e.g., a deuteron with spin-1. $q$ denotes a quark which corresponds to a jet of hadrons observed in experiments. In this paper, we consider the case of the electron scattering off a spin-1 target. This gives us the opportunity to access the tensor polarization effects in the lepton-hadron/ion scattering process. We consider the CC interaction at the tree level of electroweak theory, i.e., the exchange of a $W^\pm$ boson with momentum $q=l-l'$ between the electron and the target.
The standard variables for SIDIS used in this paper are
\begin{align}
Q^2 = -q^2,\  x=\frac{Q^2}{2 p\cdot q}, \  y=\frac{p\cdot q}{p \cdot l},\  s=(p+l)^2.
\label{eq:SIDIS-var}
\end{align}

The differential cross-section is given by
\begin{align}
  d\sigma = \frac{\alpha_{\rm em}^2}{sQ^4}\chi_W L_{\mu\nu}(l,\lambda_l, l^\prime)W^{\mu\nu}(q,p,S,k^\prime)\frac{d^3 l^\prime d^3 k^\prime}{(2\pi)^32E_{l^\prime} E_{k^\prime}}, \label{f:crosssec}
\end{align}
where $\alpha_{\rm em}$ is the fine structure constant and kinematic factor $\chi_W$ is defined as
\begin{align}
& \chi_W = \frac{Q^4}{4\left[(Q^2+M_W^2)^2 + \Gamma_W^2 M_W^2 \right] \sin^4 \theta_W}.
\end{align}
Here $ M_W, \Gamma_W$ denote the mass and width of the $W^\pm$ boson, $\theta_W$ is the Weinberg angle.
The leptonic tensor in Eq. (\ref{f:crosssec}) is defined as
\begin{align}
 &L_{\mu\nu}(l,\lambda_l, l^\prime)= (1-\lambda_l)^2 L^{\gamma\gamma}_{\mu\nu}(l,\lambda_l, l^\prime), \label{f:leptonic}
\end{align}
where
\begin{align}
   L^{\gamma\gamma}_{\mu\nu}(l,\lambda_l, l^\prime)=2\left[ l_\mu l^\prime_\nu + l_\nu l^\prime_\mu - (l\cdot l^\prime)g_{\mu\nu}+2i \lambda_l\varepsilon_{\mu\nu l l^\prime} \right].
\end{align}
We note that in Eq. (\ref{f:leptonic}) the factor $(1-\lambda_l)^2$ for negatively charged leptons $l^-$ should be changed to $(1+\lambda_l)^2$ for positively charged leptons $l^+$.
The corresponding hadronic tensors in Eq. (\ref{f:crosssec}) is given by
\begin{align}
  W^{\mu\nu}&(q,p,S,k^\prime) = \sum_X (2\pi)^3 \delta^4(p + q - k^\prime - p_X)\nonumber\\
 &\quad \times\langle p,S | J_W^\mu(0)|k^\prime;X\rangle \langle k^\prime;X | J_{W}^\nu(0) | p,S \rangle ,
\end{align}
where the current $ J_{W}^\mu(0) =\bar\psi(0) \Gamma^\mu_q \psi(0)$ with $\Gamma^\mu_q = \gamma^\mu(1- \gamma_5)$.
$W^{\mu\nu}(q,p,S,k^\prime)$ is related to the hadronic tensor $W^{\mu\nu(in)}(q,p,S)$ for the inclusive process $l N \rightarrow l' X$ by
\begin{align}
W^{\mu\nu(in)}(q,p,S) =\int \frac{d^3 k^\prime}{(2\pi)^32E_{k^\prime}}W^{\mu\nu}(q,p,S,k^\prime).
\end{align}
The superscript $(in)$ denotes the $inclusive$. It is convenient to consider the  $k_\perp^\prime$-dependent cross section in the jet production SIDIS process, i.e.,
\begin{align}
  d\sigma = \frac{\alpha_{\rm{em}}^2}{sQ^4}\chi_W L_{\mu\nu}(l,\lambda_l, l^\prime)W^{\mu\nu}(q,p,S,k_\perp^\prime) \frac{d^3 l^\prime d^2 k_\perp^\prime}{E_{l^\prime}}, \label{f:crosssection}
\end{align}
where the $k^{\prime}_z$-integrated TMD semi-inclusive hadronic tensor is given by
\begin{align}
W^{\mu\nu}(q,p,S,k_\perp^\prime) = \int \frac{dk_z^\prime}{(2\pi)^3 2E_{k^\prime}} W^{\mu\nu}(q,p,S,k^\prime).
\end{align}
In terms of the variables shown in Eq. (\ref{eq:SIDIS-var}), we have
$\frac{d^3 l^\prime}{2E_{l^\prime}} \approx \frac{y s}{4} dx d y d\psi$,
where $\psi$ is the azimuthal angle of $\vec l^\prime$ around $\vec l$, the cross section can be rewritten as an explicit form
\begin{align}
\frac{d\sigma}{dx dy d\psi d^2 k_\perp^\prime} = \frac{y \alpha_{\rm em}^2}{2 Q^4} \chi_{W}L_{\mu\nu}(l,\lambda_l, l^\prime) W^{\mu\nu}(q,p,S,k_\perp^\prime). \label{eq:Xsec-PartonModel}
\end{align}

\subsection{The general form of the cross section in terms of structure functions}

Because the hadronic tensor can not be calculated with perturbative theory, we present a general decomposition of it.
We first divide the hadronic tensor into a symmetric and an antisymmetric part, $W^{\mu\nu} = W^{S\mu\nu}+ iW^{A\mu\nu}$. Considering the weak interaction does not conserve parity, we furthermore have
\begin{align}
   W^{S\mu\nu} &=\sum_{\sigma,j} W_{\sigma j}^S h_{\sigma j}^{S\mu\nu} + \sum_{\sigma, j} \tilde W_{\sigma j}^S \tilde h_{\sigma j}^{S\mu\nu},\label{f:Wsmunu}\\
   W^{A\mu\nu} &=\sum_{\sigma,j} W_{\sigma j}^A h_{\sigma j}^{A\mu\nu} + \sum_{\sigma, j} \tilde W_{\sigma j}^A \tilde h_{\sigma j}^{A\mu\nu},\label{f:Wamunu}
\end{align}
where $h^{\mu\nu}_{\sigma j}$'s and $\tilde h^{\mu\nu}_{\sigma j}$'s represent the space reflection even and odd basic Lorentz tensors (BLTs), respectively. The subscript $\sigma$ specifies the polarizations.

The detailed discussions about the description of polarizations for spin-1 hadron and the construction of BLTs can be seen in ref. \cite{Chen:2016moq}. We do not present the discussions in this paper. However, we repeat the results here for completeness. There are 9 unpolarized BLTs given by
\begin{align}
   h^{S\mu\nu}_{Ui}&=\Big\{g^{\mu\nu}-\frac{q^\mu q^\nu}{q^2}, ~p_q^\mu  p_q^\nu,~ k_{q}^{\prime\mu}  k_{q}^{\prime\nu},  ~ p_q^{\{\mu} k_{ q}^{\prime\nu\}}\Big\},\label{f:hsU} \\
   \tilde h^{S\mu\nu}_{Ui}&=\Big\{\varepsilon^{\{\mu q p k^\prime } p_q^{\nu\}},~ \varepsilon^{\{\mu q p k^\prime }k_{ q}^{\prime\nu\}}\Big\}, \label{f:thsU} \\
   h^{A\mu\nu}_{U}&=\Big\{p_q^{[\mu} k_{q}^{\prime\nu]}\Big\},\label{f:haU} \\
   \tilde h^{A\mu\nu}_{Ui}&=\Big\{\varepsilon ^{\mu\nu qp},~ \varepsilon ^{\mu\nu qk^\prime }\Big\}.\label{f:thaU}
\end{align}
The subscript $U$ denotes the unpolarized part, and $p_q\equiv p - q(p\cdot q)/q^2$ satisfies $p_q \cdot q$ = 0.
We have also used notations $A^{\{\mu}B^{\nu\}} \equiv A^\mu B^\nu +A^\nu B^\mu$, and $A^{[\mu}B^{\nu]} \equiv A^\mu B^\nu -A^\nu B^\mu$.

The vector polarization dependent BLTs can be constructed from the unpolarized BLTs and be written as a unified form given by
\begin{align}
  h^{S\mu\nu}_{Vi} &=\Big\{[\lambda_h, (k^\prime _\perp\cdot S_T)]\tilde h^{S\mu\nu}_{Ui}, ~\varepsilon_\perp^{k^\prime S} h^{S\mu\nu}_{Uj}\Big\}, \label{f:hsV}\\
 \tilde h^{S\mu\nu}_{Vi} &=\Big\{[\lambda_h, (k^\prime _\perp\cdot S_T)]h^{S\mu\nu}_{Ui}, ~ \varepsilon_\perp^{k^\prime S} \tilde h^{S\mu\nu}_{Uj} \Big\}, \label{f:thsV}\\
  h^{A\mu\nu}_{Vi} &=\Big\{[\lambda_h, (k^\prime _\perp\cdot S_T)]\tilde h^{A\mu\nu}_{Ui},~ \varepsilon_\perp^{k^\prime S}  h^{A\mu\nu}_{U}\Big\},\label{f:haV}\\
  \tilde h^{A\mu\nu}_{Vi} &=\Big\{[\lambda_h, (k^\prime _\perp\cdot S_T)]h^{A\mu\nu}_{U},~ \varepsilon_\perp^{k^\prime S}  \tilde h^{A\mu\nu}_{Uj}\Big\},\label{f:thaV}
\end{align}
where~$\varepsilon_\perp^{k^\prime S}=\varepsilon_\perp^{\alpha\beta} k^\prime _{\perp\alpha} S_{T\beta}$, $\varepsilon_\perp^{\alpha\beta}=\varepsilon^{\mu\nu\alpha\beta}\bar n_\mu n_\nu$; $\lambda_h$ is the hadron helicity while $S_T$ is the transverse polarization component. There are 27 vector polarized BLTs in total.

The tensor polarized part is composed of $S_{LL}$-, $S_{LT}$- and $S_{TT}$-dependent parts. There are 9 $S_{LL}$-dependent BLTs, they are given by
\begin{align}
&h_{LLi}^{S\mu\nu}=S_{LL} h^{S\mu\nu}_{Ui}, &&\tilde h_{LLi}^{S\mu\nu}=S_{LL} \tilde h^{S\mu\nu}_{Ui}, \label{f:hsLL} \\
&h_{LL}^{A\mu\nu}=S_{LL} h^{A\mu\nu}_{U}, &&\tilde h_{LLi}^{A\mu\nu}=S_{LL} \tilde h^{A\mu\nu}_{Ui}. \label{f:thsLL}
\end{align}
For the $S_{LT}$ part, we have
\begin{align}
 & h^{S\mu\nu}_{LTi} =\Big\{ (k^\prime _\perp\cdot S_{LT}) h^{S\mu\nu}_{Ui}, ~\varepsilon_\perp^{k^\prime S_{LT}} \tilde h^{S\mu\nu}_{Uj}\Big\}, \label{f:hsLT}\\
 & \tilde h^{S\mu\nu}_{LTi} =\Big\{(k^\prime _\perp\cdot S_{LT})\tilde h^{S\mu\nu}_{Ui}, ~ \varepsilon_\perp^{k^\prime S_{LT}} h^{S\mu\nu}_{Uj} \Big\}, \label{f:thsLT}\\
 & h^{A\mu\nu}_{LTi} =\Big\{(k^\prime _\perp\cdot S_{LT}) h^{A\mu\nu}_{U},~ \varepsilon_\perp^{k^\prime S_{LT}}  \tilde h^{A\mu\nu}_{Uj}\Big\},\label{f:haLT}\\
 & \tilde h^{A\mu\nu}_{LTi} =\Big\{(k^\prime _\perp\cdot S_{LT})\tilde h^{A\mu\nu}_{Ui},~ \varepsilon_\perp^{k^\prime S_{LT}} h^{A\mu\nu}_{U}\Big\}. \label{f:thaLT}
\end{align}
For the $S_{TT}$ part, we have
\begin{align}
 & h^{S\mu\nu}_{TTi} =\Big\{ S_{TT}^{k^\prime k^\prime} h^{S\mu\nu}_{Ui}, ~\tilde S_{TT}^{k^\prime k^\prime}  \tilde h^{S\mu\nu}_{Uj}\Big\}, \label{f:hsTT}\\
 & \tilde h^{S\mu\nu}_{TTi} =\Big\{S_{TT}^{k^\prime k^\prime} \tilde h^{S\mu\nu}_{Ui}, ~ \tilde S_{TT}^{ k^\prime k^\prime} h^{S\mu\nu}_{Uj} \Big\}, \label{f:thsTT}\\
 & h^{A\mu\nu}_{TTi} =\Big\{S_{TT}^{k^\prime k^\prime}  h^{A\mu\nu}_{U},~ \tilde S_{TT}^{k^\prime k^\prime }  \tilde h^{A\mu\nu}_{Uj}\Big\}, \label{f:haTT}\\
 & \tilde h^{A,\mu\nu}_{TTi} =\Big\{S_{TT}^{k^\prime k^\prime} \tilde h^{A\mu\nu}_{Ui},~ \tilde S_{TT}^{ k^\prime k^\prime } h^{A\mu\nu}_{U}\Big\},\label{f:thaTT}
\end{align}
where~ $S_{TT}^{k^\prime k^\prime}=k^\prime _{\perp\alpha} S_{TT}^{\alpha\beta} k^\prime_{\perp\beta}$, $\tilde k_{\perp}^{\prime\alpha}=\varepsilon_{\perp}^{\alpha\beta} k^\prime_{\perp\beta}$.
There are 81 such BLTs in total. 

In order to calculate the cross section, we choose a coordinate system so that the momenta related to this SIDIS process take the following forms:
\begin{align}
& p^\mu = \left(p^+,0,\vec 0_\perp \right), \nonumber\\
& l^\mu = \left( \frac{1-y}{y}xp^+, \frac{Q^2}{2xyp^+}, \frac{Q\sqrt{1-y}}{y},0 \right),\nonumber\\
& q^\mu = \left( -xp^+, \frac{Q^2}{2xp^+}, \vec 0_\perp \right), \nonumber\\
& k_\perp^{\prime\mu} = k_\perp^\mu = |\vec k_\perp| \left( 0,0, \cos\varphi, \sin\varphi \right),
\end{align}
where $k_\perp^\mu$ is the quark transverse momentum in the target hadron.
The transverse vector polarization is parameterized as
\begin{align}
& S_T^\mu = |\vec S_T| \left( 0,0, \cos\varphi_S, \sin\varphi_S \right).
\end{align}
For the tensor polarization dependent parameters, we parameterize and define them as in ref.~\cite{Bacchetta:2000jk},
\begin{align}
& S_{LT}^x = |S_{LT}| \cos\varphi_{LT}, \\
& S_{LT}^y = |S_{LT}| \sin\varphi_{LT}, \\
& S_{TT}^{xx} = -S_{TT}^{yy} = |S_{TT}| \cos2\varphi_{TT}, \\
& S_{TT}^{xy} = S_{TT}^{yx} = |S_{TT}| \sin2\varphi_{TT}, \\
& |S_{LT}| = \sqrt{(S_{LT}^x)^2 + (S_{LT}^y)^2}, \\
& |S_{TT}| = \sqrt{(S_{TT}^{xx})^2 + (S_{TT}^{xy})^2}.
\end{align}

Substituting the BLTs expressed in Eqs. (\ref{f:hsU})-(\ref{f:thaTT}) into Eqs. (\ref{f:Wsmunu})-(\ref{f:Wamunu}) and making Lorentz contractions with the leptonic tensor, we obtain the general form for the cross section which is given by
\begin{align}
  \frac{d\sigma}{dxdyd\psi d^2k^\prime _\perp}&=\frac{\alpha_{\rm em}^2}{yQ^2}\chi_W \Big[ \mathcal{W}_{U}+ \lambda_h\mathcal{W}_{L} +S_{LL}\mathcal{W}_{LL} \nonumber\\
  &+|S_T|\mathcal{W}_{T} +|S_{LT}|\mathcal{W}_{LT} + |S_{TT}|\mathcal{W}_{TT} \Big]. \label{f:crossW}
\end{align}
The explicit results for the cross section in terms of structure functions for different polarization configurations are:

\begin{align}
{\cal W}_{U} &= A(y) W_{U}^T + E(y) W_{U}^L + B(y)\left( \sin\varphi \tilde W_{U1}^{\sin\varphi} + \cos\varphi W_{U1}^{\cos\varphi} \right) + E(y)\left( \sin2\varphi \tilde W_{U}^{\sin2\varphi} + \cos2\varphi W_{U}^{\cos2\varphi} \right)  \nonumber\\
& + C(y) W_{U}+ D(y)\left( \sin\varphi \tilde W_{U2}^{\sin\varphi} + \cos\varphi W_{U2}^{\cos\varphi} \right), \label{f:Wuu} \\
{\cal W}_{L} &= A(y) \tilde W_{L}^T + E(y) \tilde W_{L}^L + B(y)\left( \sin\varphi W_{L1}^{\sin\varphi} + \cos\varphi \tilde W_{L1}^{\cos\varphi} \right) + E(y)\left( \sin2\varphi W_{L}^{\sin2\varphi} + \cos2\varphi \tilde W_{L}^{\cos2\varphi} \right)  \nonumber\\
& + C(y) \tilde W_{L} + D(y) \left( \sin\varphi W_{L2}^{\sin\varphi} + \cos\varphi \tilde W_{L2}^{\cos\varphi} \right), \\
{\cal W}_{LL} &= A(y) W_{LL}^T + E(y) W_{LL}^L + B(y)\left( \sin\varphi \tilde W_{LL1}^{\sin\varphi} + \cos\varphi W_{LL1}^{\cos\varphi} \right) + E(y)\left( \sin2\varphi \tilde W_{LL}^{\sin2\varphi} + \cos2\varphi W_{LL}^{\cos2\varphi} \right)  \nonumber\\
& + C(y) W_{LL} + D(y) \left( \sin\varphi \tilde W_{LL2}^{\sin\varphi} + \cos\varphi W_{LL2}^{\cos\varphi} \right), \\
{\cal W}_{T} &= \sin\varphi_S \left[ B(y) W_{T1}^{\sin\varphi_S} + D(y) W_{T2}^{\sin\varphi_S} \right]
+ \sin(\varphi+\varphi_S) E(y) W_{T}^{\sin(\varphi+\varphi_S)} \nonumber\\
&+ \sin(\varphi-\varphi_S) \left[ A(y) W_{T}^{T,\sin(\varphi-\varphi_S)} + E(y) W_{T}^{L,\sin(\varphi-\varphi_S)} + C(y) W_{T}^{\sin(\varphi-\varphi_S)} \right] \nonumber\\
&+ \sin(2\varphi-\varphi_S) \left[ B(y) W_{T1}^{\sin(2\varphi-\varphi_S)} + D(y) W_{T2}^{\sin(2\varphi-\varphi_S)} \right]
+ \sin(3\varphi-\varphi_S) E(y) W_{T}^{\sin(3\varphi-\varphi_S)} \nonumber\\
&+ \cos\varphi_S \left[ B(y) \tilde W_{T1}^{\cos\varphi_S} + D(y) \tilde W_{T2}^{\cos\varphi_S} \right]
+ \cos(\varphi+\varphi_S) E(y) \tilde W_{T}^{\cos(\varphi+\varphi_S)} \nonumber\\
&+ \cos(\varphi-\varphi_S) \left[ A(y) \tilde W_{T}^{T,\cos(\varphi-\varphi_S)} + E(y) \tilde W_{T}^{L,\cos(\varphi-\varphi_S)} + C(y) \tilde W_{T}^{\cos(\varphi-\varphi_S)} \right] \nonumber\\
&+ \cos(2\varphi-\varphi_S) \left[ B(y) \tilde W_{T1}^{\cos(2\varphi-\varphi_S)} + D(y) \tilde W_{T2}^{\cos(2\varphi-\varphi_S)} \right]
+ \cos(3\varphi-\varphi_S) E(y) \tilde W_{T}^{\cos(3\varphi-\varphi_S)}, \\
{\cal W}_{LT} &= \sin\varphi_{LT} \left[ B(y) \tilde W_{LT1}^{\sin\varphi_{LT}} + D(y) \tilde W_{LT2}^{\sin\varphi_{LT}} \right]
+ \sin(\varphi+\varphi_{LT}) E(y) \tilde W_{LT}^{\sin(\varphi+\varphi_{LT})} \nonumber\\
&+ \sin(\varphi-\varphi_{LT}) \left[ A(y) \tilde W_{LT}^{T,\sin(\varphi-\varphi_{LT})} + E(y) \tilde W_{LT}^{L,\sin(\varphi-\varphi_{LT})} + C(y) \tilde W_{LT}^{\sin(\varphi-\varphi_{LT})} \right] \nonumber\\
&+ \sin(2\varphi-\varphi_{LT}) \left[ B(y) \tilde W_{LT1}^{\sin(2\varphi-\varphi_{LT})} + D(y) \tilde W_{LT2}^{\sin(2\varphi-\varphi_{LT})} \right]
+ \sin(3\varphi-\varphi_{LT}) E(y) \tilde W_{LT}^{\sin(3\varphi-\varphi_{LT})} \nonumber\\
&+ \cos\varphi_{LT} \left[ B(y) W_{LT1}^{\cos\varphi_{LT}} + D(y) W_{U,LT2}^{\cos\varphi_{LT}} \right]
+ \cos(\varphi+\varphi_{LT}) E(y) W_{LT}^{\cos(\varphi+\varphi_{LT})} \nonumber\\
&+ \cos(\varphi-\varphi_{LT}) \left[ A(y) W_{LT}^{T,\cos(\varphi-\varphi_{LT})} + E(y) W_{LT}^{L,\cos(\varphi-\varphi_{LT})} + C(y) W_{LT}^{\cos(\varphi-\varphi_{LT})} \right] \nonumber\\
&+ \cos(2\varphi-\varphi_{LT}) \left[ B(y) W_{LT1}^{\cos(2\varphi-\varphi_{LT})} + D(y) W_{LT2}^{\cos(2\varphi-\varphi_{LT})} \right]
+ \cos(3\varphi-\varphi_{LT}) E(y) W_{LT}^{\cos(3\varphi-\varphi_{LT})}, \\
{\cal W}_{TT} &= \sin(\varphi-2\varphi_{TT}) \left[ B(y) \tilde W_{TT1}^{\sin(\varphi-2\varphi_{TT})} + D(y) \tilde W_{TT2}^{\sin(\varphi-2\varphi_{TT})} \right]
+ \sin2\varphi_{TT} E(y) \tilde W_{TT}^{\sin2\varphi_{TT}} \nonumber\\
&+ \sin(2\varphi-2\varphi_{TT}) \left[ A(y) \tilde W_{TT}^{T,\sin(2\varphi-2\varphi_{TT})} + E(y) \tilde W_{TT}^{L,\sin(2\varphi-2\varphi_{TT})} + C(y) \tilde W_{TT}^{\sin(2\varphi-2\varphi_{TT})} \right] \nonumber\\
&+ \sin(3\varphi-2\varphi_{TT}) \left[ B(y) \tilde W_{TT1}^{\sin(3\varphi-2\varphi_{TT})} + D(y) \tilde W_{TT2}^{\sin(3\varphi-2\varphi_{TT})} \right]
+ \sin(4\varphi-2\varphi_{TT}) E(y) \tilde W_{TT}^{\sin(4\varphi-2\varphi_{TT})} \nonumber\\
&+ \cos(\varphi-2\varphi_{TT}) \left[ B(y) W_{TT1}^{\cos(\varphi-2\varphi_{TT})} + D(y) W_{TT2}^{\cos(\varphi-2\varphi_{TT})} \right]
+ \cos2\varphi_{TT} E(y) W_{TT}^{\cos2\varphi_{TT}} \nonumber\\
&+ \cos(2\varphi-2\varphi_{TT}) \left[ A(y) W_{TT}^{T,\cos(2\varphi-2\varphi_{TT})} + E(y) W_{TT}^{L,\cos(2\varphi-2\varphi_{TT})} + C(y) W_{TT}^{\cos(2\varphi-2\varphi_{TT})} \right] \nonumber\\
&+ \cos(3\varphi-2\varphi_{TT}) \left[ B(y) W_{TT1}^{\cos(3\varphi-2\varphi_{TT})} + D(y) W_{TT2}^{\cos(3\varphi-2\varphi_{TT})} \right]
+ \cos(4\varphi-2\varphi_{TT}) E(y) W_{TT}^{\cos(4\varphi-2\varphi_{TT})}. \label{f:WLTT}
\end{align}

Here we have defined the following functions of $y$,
\begin{align}
& A(y) = y^2-2y+2, \nonumber\\
& B(y) = 2(2-y)\sqrt{1-y}, \nonumber\\
& C(y) = y(2-y), \nonumber\\
& D(y) = 2y\sqrt{1-y}, \nonumber\\
& E(y) = 2(1-y).
\end{align}
There exist 81 structure functions which correspond to the BLTs. However, not all of them contribute in the charged current SIDIS process at twist-3 level. It will be clear in Sec. \ref{sec:result} where we present the structure functions in terms of the gauge invariant PDFs.

\section{The hadronic tensor in the QCD parton model} \label{sec:partonmodel}
\subsection{The collinear expansion}
\begin{figure} [b]
\centering
\includegraphics[width= 0.95\linewidth]{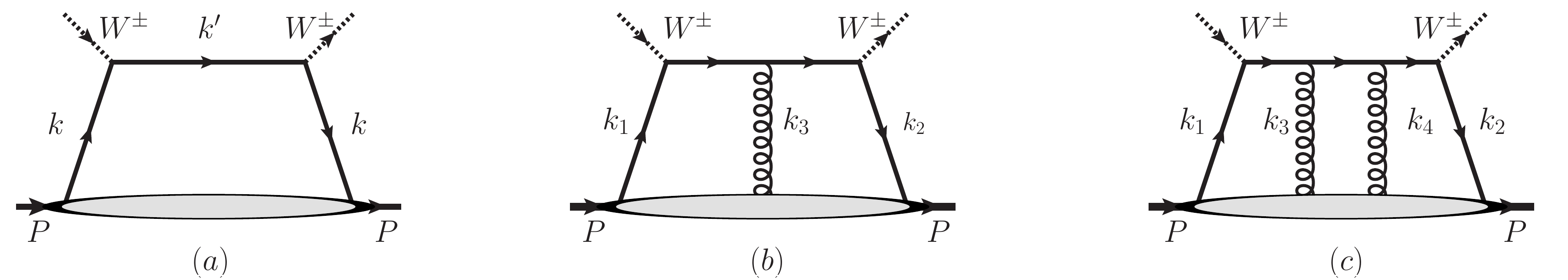}
\caption{The first few diagrams of the Feynman diagram series with exchange of $j$ gluons, where $j=0,~1$ and $2$ for diagrams $(a)$, $(b)$ and $(c)$ respectively.}
\label{CC}
\end{figure}

In the parton model, the hadronic tensor can be expressed in terms of gauge-invariant TMD PDFs.
At the leading order twist-3 level, we need to consider the contributions from the series of diagrams shown in Fig.~\ref{CC}, i.e., the multiple gluon scattering contributions.

After collinear expansion, the hadronic tensor is expressed in terms of
the gauge-invariant quark-quark and quark-gluon-quark correlators and calculable hard parts \cite{Liang:2006wp,Song:2010pf,Song:2013sja},
\begin{align}
W_{\mu\nu} (q,p,S,k^\prime) = \sum_{j,c} \tilde W_{\mu\nu}^{(j,c)} (q,p,S,k^\prime),
\end{align}
where $j$ denotes the number of gluons exchanged and $c$ denotes different cuts. After integration over $k_z^\prime$, $\tilde W_{r,\mu\nu}^{(j,c)}$'s are simplified as
\begin{align}
& \tilde{W}_{\mu\nu}^{(0)}(q,p,S,k_\perp^\prime) = \frac{1}{2}{\rm Tr}\left[\hat{h}_{\mu \nu}^{(0)} \hat{\Phi}^{(0)}(x,k_\perp)\right] , \label{f:W0munu}\\
& \tilde{W}_{\mu\nu}^{(1, L)}(q,p,S,k_\perp^\prime) = \frac{1}{4 p \cdot q}{\rm Tr}\left[\hat{h}_{\mu\nu}^{(1) \rho} \hat{\varphi}_{\rho}^{(1)}(x,k_\perp)\right], \label{f:W1Lmunu}
\end{align}
up to the twist-3 level. The hard parts $h$'s are
\begin{align}
& \hat h_{\mu\nu}^{(0)} = \Gamma_{\mu}^q \slashed n \Gamma_{\nu}^q / p^+, \qquad \hat{h}_{\mu\nu}^{(1) \rho}=\Gamma_{\mu}^q \slashed{\bar n} \gamma_{\perp}^{\rho} \slashed n \Gamma_{\nu}^q .
\end{align}
The corresponding gauge-invariant quark-quark and quark-gluon-quark correlators are defined as
\begin{align}
  \hat{\Phi}^{(0)}\left(x, k_{\perp}\right) =& \int \frac{p^{+} d y^{-} d^{2} y_{\perp}}{(2 \pi)^{3}} e^{i x p^{+} y^{-}-i \vec{k}_{\perp} \cdot \vec{y}_{\perp}} \nonumber\\
  &\times \langle N|\bar{\psi}(0) {\cal L}(0, y) \psi(y)| N\rangle, \\
  \hat{\varphi}_{\rho}^{(1)}\left(x, k_{\perp}\right) =& \int \frac{p^{+} d y^{-} d^{2} y_{\perp}}{(2 \pi)^{3}} e^{i x p^+ y^- - i \vec{k}_\perp \cdot \vec{y}_\perp}\nonumber\\
  &\times  \langle N | \bar{\psi}(0) D_{\perp \rho}(0) {\cal L}(0, y) \psi(y)| N\rangle,
\end{align}
where $D_\rho(y) = -i\partial_\rho + g A_\rho(y)$ is the covariant derivative.
${\cal L}(0, y)$ is the gauge link obtained from the collinear expansion procedure.

\subsection{Decomposition of the quark-quark and quark-gluon-quark correlators}

The quark-quark and quark-gluon-quark correlators are $4\times 4$ matrices in Dirac space which can be decomposed in terms of the gamma matrices, $\{I, i\gamma^5, \gamma^\alpha, \gamma^\alpha\gamma^5, i\sigma^{\alpha\beta}\gamma^5\}$, and the corresponding coefficient functions.
However, in the jet production SIDIS process where the final hadron is not considered only the chiral even PDFs are involved since chirality does not flip. Thus we only need to consider the $\gamma^\alpha$- and the $\gamma^\alpha\gamma^5$-terms in the decomposition of these correlators. We have
\begin{align}
& \hat \Phi^{(0)}= \frac{1}{2}\left[\gamma^\alpha \Phi^{(0)}_\alpha + \gamma^\alpha\gamma_5 \tilde\Phi^{(0)}_\alpha \right], \\
& \hat \varphi_\rho^{(1)}= \frac{1}{2}\left[\gamma^\alpha \varphi_{\rho\alpha}^{(1)} + \gamma^\alpha\gamma_5 \tilde\varphi_{\rho\alpha}^{(1)}\right] .
\end{align}

The TMD PDFs can be obtained through the decomposition of the coefficient functions, $\Phi^{(0)}_\alpha, \tilde\Phi^{(0)}_\alpha, \varphi_{\rho\alpha}^{(1)}$ and $\tilde\varphi_{\rho\alpha}^{(1)}$.
Following the convention in ref.~\cite{Wei:2016far}, we write down the complete decompositions of the coefficient functions at twist-3 level for spin-1 particles.
\begin{align}
  \Phi^{(0)}_\alpha &=p^+ \bar n_\alpha\Bigl(f_1+S_{LL}f_{1LL}-\frac{k_\perp \cdot \tilde S_T}{M}f^\perp_{1T} \nonumber\\ &+\frac{k_\perp \cdot S_{LT}}{M}f_{1LT}^{\perp}+\frac{S_{TT}^{kk}}{M^2}f_{1TT}^{\perp} \Bigr) +k_{\perp\alpha}\Big( f^\perp +S_{LL}f_{LL}^\perp  \Big)  \nonumber\\
  &- M\tilde S_{T\alpha}f_T + M S_{LT\alpha}f_{LT}+ S^k_{TT\alpha} f_{TT} - \lambda_h \tilde k_{\perp\alpha} f_L^\perp \nonumber\\ & -\frac{k_{\perp\langle\alpha}k_{\perp\beta\rangle}}{M} \Bigl( \tilde S_T^\beta f_T^\perp + S_{LT}^\beta f_{LT}^\perp + \frac{ S_{TT}^{k\beta}}{M} f_{TT}^\perp \Bigr),
\label{eq:Xi0Peven}\\
  \tilde\Phi^{(0)}_\alpha &=p^+\bar n_\alpha\Bigl(-\lambda_hg_{1L}+\frac{k_\perp\cdot S_T}{M}g^\perp_{1T}\nonumber\\
  &+ \frac{k_\perp \cdot \tilde S_{LT}}{M} g_{1LT}^\perp - \frac{\tilde S_{TT}^{k k}}{M^2} g_{1TT}^\perp\Bigr)-\tilde k_{\perp\alpha} \Big( g^\perp+S_{LL}g_{LL}^\perp \Big) \nonumber\\
  &- M S_{T\alpha}g_T - M \tilde S_{LT\alpha}g_{LT} - \tilde S^{k}_{TT\alpha} g_{TT} -\lambda_h k_{\perp\alpha} g_L^\perp \nonumber\\
  &+ \frac{k_{\perp\langle\alpha}k_{\perp\beta\rangle}}{M} \Bigl( S_T^\beta g_T^\perp - \tilde S_{LT}^\beta g_{LT}^\perp - \frac{ S_{TT}^{k\beta}}{M} \tilde g_{TT}^\perp \Bigr).
\label{eq:Xi0Podd}
\end{align}
 For the quark-gluon-quark correlator, we have
\begin{align}
  \varphi^{(1)}_{\rho\alpha}&=p^+\bar n_\alpha\Biggl[k_{\perp\rho}\big( f^\perp_d+ S_{LL}f_{dLL}^\perp\big)- M\tilde S_{T\rho}f_{dT} \nonumber\\
  & +MS_{LT\rho} f_{dLT}+S_{TT\rho}^k f_{dTT} -\lambda_h \tilde k_{\perp\rho} f_{dL}^\perp \nonumber\\
  & -\frac{k_{\perp\langle\rho}k_{\perp\beta\rangle}}{M} \Bigl( \tilde S_T^\beta f_{dT}^\perp + S_{LT}^\beta f_{dLT}^\perp + \frac{S_{TT}^{k\beta}}{M} f_{dTT}^\perp \Bigr)\Biggr], \label{eq:Xi1Peven} \\
  \tilde \varphi^{(1)}_{\rho\alpha}&=ip^+\bar n_\alpha\Biggl[\tilde k_{\perp\rho}\big( g^\perp_d+ S_{LL}g_{dLL}^\perp\big) + MS_{T\rho}g_{dT}\nonumber\\
  & + M\tilde S_{LT\rho} g_{dLT} + \tilde S_{TT\rho}^{k} g_{dTT}+\lambda_h k_{\perp\rho} g_{dL}^\perp  \nonumber\\
  & - \frac{k_{\perp\langle\rho}k_{\perp\beta\rangle}}{M} \Bigl( S_T^\beta g_{dT}^\perp - \tilde S_{LT}^\beta g_{dLT}^\perp - \frac{\tilde S_{TT}^{k\beta}}{M} g_{dTT}^\perp \Bigr) \Biggr],
\label{eq:Xi1Podd}
\end{align}
where
$S_{TT}^{k\beta} \equiv S_{TT}^{\alpha\beta}k_{\perp\alpha}$, and $\tilde S_{TT}^{k\beta} \equiv \varepsilon_{\perp\mu}^{\beta} S_{TT}^{k\mu}$. We have $\frac{1}{M} S_{TT}^{k\beta}$ behaves as a Lorentz vector like $S_{LT}^\beta$, $\frac{1}{M} \tilde S_{TT}^{k\beta}$ and $\tilde S_{LT}^\beta$ behave as axial vectors like $S_{T}^\beta$.
Not all of these TMD PDFs shown in Eqs.~(\ref{eq:Xi0Peven})-(\ref{eq:Xi1Podd}) are independent. We use the QCD equation of motion to obtain the following equations to eliminate PDFs which are not independent, i.e.,
\begin{align}
 x p^{+} \Phi^{(0) \rho} &=-g_{\perp}^{\rho \sigma} \operatorname{Re} \varphi_{\sigma+}^{(1)}-\varepsilon_{\perp}^{\rho \sigma} \operatorname{Im} \tilde{\varphi}_{\sigma+}^{(1)}, \label{eq:eom1}\\
 x p^{+} \tilde{\Phi}^{(0) \rho} &=-g_{\perp}^{\rho \sigma} \operatorname{Re} \tilde{\varphi}_{\sigma+}^{(1)}-\varepsilon_{\perp}^{\rho \sigma} \operatorname{Im} \varphi_{\sigma+}^{(1)}.\label{eq:eom2}
\end{align}
By inserting Eqs.~(\ref{eq:Xi0Peven})-(\ref{eq:Xi1Podd}) into Eqs.~(\ref{eq:eom1}) and (\ref{eq:eom2}), we can get the relationships between the twist-3 TMD PDFs defined via the quark-quark correlator and those defined via the quark-gluon-quark correlator.
They can be written in a unified form, i.e.,
\begin{align}
f_{d S}^{K}-g_{d S}^{K}=-x\left(f_{S}^{K}-i g_{S}^{K}\right),\label{f:formulaEOM}
\end{align}
where $K$ denotes $\perp$ while $S$ denotes $L$, $T$, $LL$, $LT$ and $TT$ whenever applicable.

\subsection{The hadronic tensor results}

Substituting the Lorentz decomposition expressions of the correlators into the hadronic tensor and carrying out the traces we can obtain the results for the hadronic tensor up to twist-3.
The relevant traces we need are
\begin{align}
 &p^+{\rm Tr} \left[ \gamma_\alpha \hat h_{\mu\nu}^{(0)} \right] = -8 \varrho_{\mu\nu\alpha} - 8i\varepsilon_{\alpha n \mu\nu} ,\\
&p^+{\rm Tr} \left[ \gamma_\alpha \gamma_5 \hat h_{\mu\nu}^{(0)} \right] = -8 \varrho_{\mu\nu\alpha} + 8i \varepsilon_{n \alpha \mu\nu},\\
& {\rm Tr} \left[ \slashed{\bar n} \hat h_{\mu\nu}^{(1)\rho} \right] = -16 g_{\perp\nu}^\rho \bar n_\mu - 16i \varepsilon_{\perp\nu}^\rho \bar n_\mu,\\
& {\rm Tr} \left[ \slashed{\bar n} \gamma_5 \hat h_{\mu\nu}^{(1)\rho} \right] = -16 g_{\perp\nu}^\rho \bar n_\mu - 16i \varepsilon_{\perp\nu}^{\rho} \bar n_\mu.
\end{align}
Here the $\mu\nu$-symmetric tensor is defined as $\varrho_{\mu\nu\alpha} \equiv g_{\mu\nu} n_\alpha -  g_{\nu\alpha} n_\mu - g_{\mu\alpha} n_\nu$.

The hadronic tensor at the leading twist coming from the quark-quark correlator in Eqs.~(\ref{eq:Xi0Peven}) and (\ref{eq:Xi0Podd}) is given by
\begin{align}
\tilde W_{t2}^{(0)\mu\nu} =&
-2\left(g_{\perp}^{\mu\nu} + i \varepsilon_{\perp}^{\mu\nu} \right) \Bigl(f_1+S_{LL}f_{1LL}-\frac{k_\perp \cdot \tilde S_T}{M}f^\perp_{1T}\nonumber\\
& \qquad\qquad\qquad+\frac{k_\perp \cdot S_{LT}}{M}f_{1LT}^{\perp}+\frac{S_{TT}^{kk}}{M^2}f_{1TT}^{\perp} \Bigr) \nonumber\\
& - 2\left(g_{\perp}^{\mu\nu} + i\varepsilon_{\perp}^{\mu\nu} \right) \Bigl(-\lambda_hg_{1L}+\frac{k_\perp\cdot S_T}{M}g^\perp_{1T} \nonumber\\
&\qquad\qquad+ \frac{k_\perp \cdot \tilde S_{LT}}{M} g_{1LT}^\perp - \frac{\tilde S_{TT}^{k k}}{M^2} g_{1TT}^\perp\Bigr). \label{f:Wt2leadingmunu}
\end{align}
The twist-3 hadronic tensor comes also from both the quark-quark correlator and quark-gluon-quark correlators.
After using the equation of motion in Eq. (\ref{f:formulaEOM}), we get the complete hadronic tensor at twist-3 level,

\begin{align}
\frac{(p \cdot q)}{2} \tilde W_{t3}^{\mu\nu} 
&= \Bigl[k_\perp^{\{\mu} \bar q^{\nu\}} + i\tilde k_\perp^{[\mu} \bar q^{\nu]} \Bigr] \left( f^\perp + S_{LL} f_{LL}^\perp \right)- \Bigl[\tilde k_\perp^{\{\mu} \bar q^{\nu\}} - ik_\perp^{[\mu} \bar q^{\nu]} \Bigr] \lambda_h f_L^\perp - \Bigl[\tilde S_T^{\{\mu} \bar q^{\nu\}} - i S_T^{[\mu} \bar q^{\nu]} \Bigr] M f_T \nonumber\\
&+ \Bigl[S_{LT}^{\{\mu} \bar q^{\nu\}} + i\tilde S_{LT}^{[\mu} \bar q^{\nu]} \Bigr] M f_{LT} + \Bigl[S_{TT}^{k\{\mu} \bar q^{\nu\}} + i\tilde S_{TT}^{k[\mu} \bar q^{\nu]} \Bigr] f_{TT} \nonumber\\
&- \Biggl[\left( \frac{k_\perp\cdot \tilde S_T}{M} k_\perp^{\{\mu} \bar q^{\nu\}} - \frac{k_\perp^2}{2M} \tilde S_T^{\{\mu} \bar q^{\nu\}} \right) + i\left(\frac{k_\perp\cdot S_T}{M} k_\perp^{[\mu} \bar q^{\nu]} - \frac{k_\perp^2}{2M}S_T^{[\mu} \bar q^{\nu]} \right) \Biggr] f_T^\perp \nonumber\\
&- \Biggl[\left( \frac{k_\perp\cdot S_{LT}}{M} k_\perp^{\{\mu} \bar q^{\nu\}}- \frac{k_\perp^2}{2M} S_{LT}^{\{\mu} \bar q^{\nu\}} \right) + i\left(\frac{k_\perp\cdot S_{LT}}{M}\tilde k_\perp^{[\mu} \bar q^{\nu]} - \frac{k_\perp^2}{2M}\tilde S_{LT}^{[\mu} \bar q^{\nu]} \right) \Biggr] f_{LT}^\perp \nonumber\\
&- \Biggl[\left( \frac{S_{TT}^{kk}}{M^2} k_\perp^{\{\mu} \bar q^{\nu\}} - \frac{k_\perp^2}{2M^2} S_{TT}^{k\{\mu} \bar q^{\nu\}} \right) + i \left(\frac{S_{TT}^{kk}}{M^2}\tilde k_\perp^{[\mu} \bar q^{\nu]} - \frac{k_\perp^2}{2M^2}\tilde S_{TT}^{k[\mu} \bar q^{\nu]} \right) \Biggr] f_{TT}^\perp \nonumber\\
&- \Bigl[\tilde k_\perp^{\{\mu} \bar q^{\nu\}} - ik_\perp^{[\mu} \bar q^{\nu]}\Bigr] \left( g^\perp + S_{LL} g_{LL}^\perp \right)- \Bigl[k_\perp^{\{\mu} \bar q^{\nu\}} + i \tilde k_\perp^{[\mu} \bar q^{\nu]}\Bigr] \lambda_h g_L^\perp - \Bigl[S_T^{\{\mu} \bar q^{\nu\}} + i\tilde S_T^{[\mu} \bar q^{\nu]} \Bigr] M g_T \nonumber\\
&- \Bigl[\tilde S_{LT}^{\{\mu} \bar q^{\nu\}} - iS_{LT}^{[\mu} \bar q^{\nu]} \Bigr] M g_{LT} - \Bigl[\tilde S_{TT}^{k\{\mu} \bar q^{\nu\}} - iS_{TT}^{k[\mu} \bar q^{\nu]} \Bigr] g_{TT} \nonumber\\
&+ \Biggl[\left( \frac{k_\perp\cdot S_T}{M} k_\perp^{\{\mu} \bar q^{\nu\}} - \frac{k_\perp^2}{2M} S_T^{\{\mu} \bar q^{\nu\}} \right) + i\left(\frac{k_\perp\cdot S_T}{M} \tilde k_\perp^{[\mu} \bar q^{\nu]} - \frac{k_\perp^2}{2M} \tilde S_T^{[\mu} \bar q^{\nu]} \right) \Biggr] g_T^\perp \nonumber\\
&- \Biggl[\left( \frac{k_\perp\cdot \tilde S_{LT}}{M} k_\perp^{\{\mu} \bar q^{\nu\}} - \frac{k_\perp^2}{2M} \tilde S_{LT}^{\{\mu} \bar q^{\nu\}} \right) + i\left(\frac{k_\perp\cdot S_{LT}}{M} k_\perp^{[\mu} \bar q^{\nu]} - \frac{k_\perp^2}{2M} S_{LT}^{[\mu} \bar q^{\nu]} \right) \Biggr] g_{LT}^\perp \nonumber\\
&- \Biggl[\left( \frac{k_\perp\cdot \tilde S_{TT}^{k}}{M^2} k_\perp^{\{\mu} \bar q^{\nu\}}- \frac{k_\perp^2}{2M^2} \tilde S_{TT}^{k\{\mu} \bar q^{\nu\}} \right) + i\left(\frac{k_\perp\cdot S_{TT}^{k}}{M^2} k_\perp^{[\mu} \bar q^{\nu]} - \frac{k_\perp^2}{2M^2} S_{TT}^{k[\mu} \bar q^{\nu]} \right) \Biggr] g_{TT}^\perp, \label{f:Wt3munu}
\end{align}
where $\bar q^\mu = q^\mu + 2xp^\mu$.
From $q\cdot\bar q = q\cdot k_\perp = 0$ and $q\cdot S_T = q\cdot S_{LT} = q\cdot S_{TT}^{k}/M = 0$, we see clearly that the full twist-3 hadronic tensor satisfies current conservation, $q_\mu \tilde W^{\mu\nu}_{t3} = q_\nu \tilde W^{\mu\nu}_{t3} = 0$.

\section{The cross section up to twist-3} \label{sec:crosssection}

Substituting the leading twist hadronic tensor the leptonic tensor into Eq.~(\ref{f:crosssection}) yields the leading twist cross section.
\begin{align}
  &\frac{d\sigma_{t2}}{dx dy d\psi d^2 k_\perp} =\frac{\alpha_{\rm em}^2 }{y Q^2}8\chi_WT_0(y)\bigg\{ (f_1+S_{LL}f_{1LL})- \lambda_h g_{1L} +|S_T|k_{\perp M}\left[\sin(\varphi-\varphi_S) f^\perp_{1T}-\cos(\varphi-\varphi_S) g^\perp_{1T}\right] \nonumber\\
  &-|S_{LT}|k_{\perp M}\left[\sin(\varphi-\varphi_{LT})g^\perp_{1LT}+\cos(\varphi-\varphi_{LT})f^\perp_{1LT}\right] -|S_{TT}|k_{\perp M}^2\left[\sin(2\varphi-2\varphi_{TT}) g^\perp_{1TT}-\cos(2\varphi-2\varphi_{TT})f^\perp_{1TT}\right] \bigg\}, \label{f:crosst2}
\end{align}
where we have defined $k_{\perp M} = |\vec k_\perp|/M$ and $T_0(y) =  A(y) +  C(y)$ to simplify the expressions. Since $k_\perp =k_\perp^\prime$ in the $\gamma^* N$ frame, in this case only $k_\perp$ is used in Eq. (\ref{f:crosst2}) and the following context.
In order to obtain Eq. (\ref{f:crosst2}) $\lambda_l =-1$ has been used for negatively charged leptons ($l^-$) scattering process. For the positively charged lepton scattering case $\lambda_l = +1$.
Similarly, substituting the twist-3 hadronic tensor and the leptonic tensor into Eq. (\ref{f:crosssection}) yields the twist-3 cross section. It is given by
\begin{align}
  \frac{d\sigma_{t3}}{dx dy d\psi d^2 k_\perp} &= -\frac{\alpha_{\rm{em}}^2 }{y Q^2}16\chi_Wx\kappa_M T_2(y) \biggl\{ k_{\perp M}\cos\varphi (f^\perp+S_{LL}f^\perp_{LL})+k_{\perp M}\sin\varphi (g^\perp+S_{LL}g^\perp_{LL}) +\lambda_h k_{\perp M}\Big[\sin\varphi f^\perp_L - \cos\varphi g_L^\perp\Big] \nonumber\\
  +&|S_T|\Big[\sin\varphi_S f_T -\cos\varphi_S g_T +\sin(2\varphi-\varphi_S) \frac{k_{\perp M}^2}{2}f^\perp_T -\cos(2\varphi-\varphi_S) \frac{k_{\perp M}^2}{2}g^\perp_T \Big] \nonumber\\
  +&|S_{LT}|\Big[\sin\varphi_{LT}g_{LT} +\cos\varphi_{LT}f_{LT}  +\sin(2\varphi-\varphi_{LT}) \frac{k_{\perp M}^2}{2} g^\perp_{LT} + \cos(2\varphi-\varphi_{LT}) \frac{k_{\perp M}^2}{2} f^\perp_{LT} \Big] \nonumber\\
  +&|S_{TT}|k_{\perp M}\Big[\sin(\varphi-2\varphi_{TT})  g_{TT} -\cos(\varphi-2\varphi_{TT}) f_{TT} -\sin(3\varphi-2\varphi_{TT})\frac{k_{\perp M}^2}{2} g^\perp_{TT} -\cos(3\varphi-2\varphi_{TT}) \frac{k_{\perp M}^2}{2} f^\perp_{TT} \Big]
 \biggr\}. \label{f:crosst3}
\end{align}

Here we have defined $\kappa_M=M/Q$ and $T_2(y) = B(y) + D(y)$ to simplify the expression. The condition, $\lambda_l =-1$, is also used here.

Here we note again that the cross sections given in Eqs. (\ref{f:crosst2})-(\ref{f:crosst3}) are present for the negatively charged lepton (electron) case. In this case, $\lambda_l =-1$. For the positively charged lepton (positron) case, $\lambda_l =+1$ and this results in that those terms related to the longitudinally polarized PDFs ($g's$) should change to the opposite signs, e.g., $- \lambda_h g_{1L}\to + \lambda_h g_{1L}$ for the leading twist contributions and $+\sin\varphi_{LT}g_{LT} \to -\sin\varphi_{LT}g_{LT}$ for the twist-3 contributions. From the cross section we see that the CC experiments provide more complementary information on the partonic structure of nucleons as they probe combinations of quark flavors different from those accessible in NC ones.


\section{Structure functions and asymmetries results up to twist-3} \label{sec:result}

In Sec. \ref{sec:formalism}, we have presented the general form of the cross section in terms of structure functions.
In the previous section we have also presented the cross section in terms of the gauge invariant TMD PDFs.
They match to each other. In this section, we present the structure functions and azimuthal asymmetries results in terms of the gauge invariant TMD PDFs. Furthermore we will calculate the charge asymmetries induced by the exchange of the incoming lepton in the lepton scattering process.

\subsection{Structure functions results}

We first present the structure functions in terms of gauge invariant PDFs. For the leading twist part, we have
\begin{align}
  & W^T_{U} = 8 f_1, \\
  & W_{U} = 8 f_1, \\
  & \tilde W^T_{L} = - 8 g_{1L}, \\
  & \tilde W_{L} = - 8 g_{1L}, \\
  & W^T_{U,LL} = 8 f_{1LL}, \\
  & W_{U,LL} = 8 f_{1LL}, \\
  & \tilde W_{T}^{T,\cos(\varphi-\varphi_S)} = - 8 k_{\perp M} g^\perp_{1T}, \\
  & \tilde W_{T}^{\cos(\varphi-\varphi_S)} = - 8 k_{\perp M} g^\perp_{1T}, \\
  & W_{T}^{T,\sin(\varphi-\varphi_S)} = 8 k_{\perp M} f^\perp_{1T}, \\
  & W_{T}^{\sin(\varphi-\varphi_S)} = 8 k_{\perp M} f^\perp_{1T}, \\
  & W_{LT}^{T,\cos(\varphi-\varphi_{LT})} = - 8 k_{\perp M} f^\perp_{1LT}, \\
  & W_{LT}^{\cos(\varphi-\varphi_{LT})} = - 8 k_{\perp M} f^\perp_{1LT}, \\
  & \tilde W_{LT}^{T,\sin(\varphi-\varphi_{LT})} = - 8 k_{\perp M} g^\perp_{1LT}, \\
  & \tilde W_{LT}^{\sin(\varphi-\varphi_{LT})} = - 8 k_{\perp M} g^\perp_{1LT}, \\
  & W_{TT}^{T,\cos(2\varphi-2\varphi_{TT})} =  8 k^2_{\perp M} f^\perp_{1TT}, \\
  & W_{TT}^{\cos(2\varphi-2\varphi_{TT})} =  8 k^2_{\perp M} f^\perp_{1TT}, \\
  & \tilde W_{TT}^{T,\sin(2\varphi-2\varphi_{TT})} = - 8 k^2_{\perp M} g^\perp_{1TT}, \\
  & \tilde W_{TT}^{\sin(2\varphi-2\varphi_{TT})} = - 8 k^2_{\perp M} g^\perp_{1TT}.
\end{align}
In total we have 18 structure functions which contribute to the leading twist. We also present the twist-3 part, we have
\begin{align}
  & W_{U1}^{\cos\varphi} = -16x\kappa_M k_{\perp M} f^\perp, \\
  & \tilde W_{U1}^{\sin\varphi} = -16x\kappa_M k_{\perp M}  g^\perp, \\
  & W_{U2}^{\cos\varphi} = -16x\kappa_M k_{\perp M} f^\perp, \\
  & \tilde W_{U2}^{\sin\varphi} = -16x\kappa_M k_{\perp M}  g^\perp, \\
  & \tilde W_{L1}^{\cos\varphi} = 16x\kappa_M k_{\perp M} g_L^\perp, \\
  & W_{L1}^{\sin\varphi} = -16x\kappa_M k_{\perp M} f_L^\perp, \\
  & \tilde W_{L2}^{\cos\varphi} = 16x\kappa_M k_{\perp M} g_L^\perp, \\
  & W_{L2}^{\sin\varphi} = -16x\kappa_M k_{\perp M} f_L^\perp, \\
  & W_{LL1}^{\cos\varphi} = -16x\kappa_M k_{\perp M} f_{LL}^\perp, \\
  & \tilde W_{LL1}^{\sin\varphi} = -16x\kappa_M k_{\perp M} g_{LL}^\perp, \\
  & W_{LL2}^{\cos\varphi} = -16x\kappa_M k_{\perp M} f_{LL}^\perp, \\
  & \tilde W_{LL2}^{\sin\varphi} = -16x\kappa_M k_{\perp M} g_{LL}^\perp, \\
  & \tilde W_{T1}^{\cos\varphi_S} = 16x\kappa_M g_T, \\
  & W_{T1}^{\sin\varphi_S} = -16x\kappa_M f_T, \\
  & \tilde W_{T2}^{\cos\varphi_S} = 16x\kappa_M g_T, \\
  & W_{T2}^{\sin\varphi_S} = -16x\kappa_M  c_3^ec_3^q f_T, \\
  & \tilde W_{T1}^{\cos(2\varphi-\varphi_S)} = 8x\kappa_M k^2_{\perp M} g_T^\perp, \\
  & W_{T1}^{\sin(2\varphi-\varphi_S)} = -8x\kappa_M k^2_{\perp M} f_T^\perp, \\
  & \tilde W_{T2}^{\cos(2\varphi-\varphi_S)} = 8x\kappa_M k^2_{\perp M}  g_T^\perp, \\
  & W_{T2}^{\sin(2\varphi-\varphi_S)} = -8x\kappa_M k^2_{\perp M} f_T^\perp, \\
  & W_{LT1}^{\cos\varphi_{LT}} = -16x\kappa_M f_{LT}, \\
  & \tilde W_{LT1}^{\sin\varphi_{LT}} = -16x\kappa_M g_{LT}, \\
  & W_{LT2}^{\cos\varphi_{LT}} = -16x\kappa_M f_{LT}, \\
  & \tilde W_{LT2}^{\sin\varphi_{LT}} = -16x\kappa_M g_{LT}, \\
  & W_{LT1}^{\cos(2\varphi-\varphi_{LT})} = -8x\kappa_M k^2_{\perp M} f_{LT}^\perp, \\
  & \tilde W_{LT1}^{\sin(2\varphi-\varphi_{LT})} = -8x\kappa_M k^2_{\perp M} g_{LT}^\perp, \\
  & W_{LT2}^{\cos(2\varphi-\varphi_{LT})} = -8x\kappa_M k^2_{\perp M} f_{LT}^\perp, \\
  & \tilde W_{LT2}^{\sin(2\varphi-\varphi_{LT})} = -8x\kappa_M k^2_{\perp M} g_{LT}^\perp, \\
& W_{TT1}^{\cos(\varphi-2\varphi_{TT})} = 16x\kappa_M k_{\perp M} f_{TT}, \\
& W_{TT2}^{\cos(\varphi-2\varphi_{TT})} = 16x\kappa_M k_{\perp M} f_{TT}, \\
& \tilde W_{TT1}^{\sin(\varphi-2\varphi_{TT})} = -16x\kappa_M k_{\perp M} g_{TT}, \\
& \tilde W_{TT2}^{\sin(\varphi-2\varphi_{TT})} = -16x\kappa_M k_{\perp M} g_{TT}, \\
& W_{TT1}^{\cos(3\varphi-2\varphi_{TT})} = 8x\kappa_M k_{\perp M}^3 f_{TT}^\perp, \\
& W_{TT2}^{\cos(3\varphi-2\varphi_{TT})} = 8x\kappa_M k_{\perp M}^3  f_{TT}^\perp, \\
& \tilde W_{TT1}^{\sin(3\varphi-2\varphi_{TT})} = 8x\kappa_M k_{\perp M}^3 g_{TT}^\perp, \\
& \tilde W_{TT2}^{\sin(3\varphi-2\varphi_{TT})} = 8x\kappa_M k_{\perp M}^3 g_{TT}^\perp.
\end{align}
In total we have 36 structure functions contribute at twist-3.

Here we only present the results with the negatively charged leptons scattering process. The reaction would choose the $e^- +U \to \nu_e +D$ channel for the CC interactions. Here $U, D$ denote the u-type and d-type PDFs,
\begin{align}
  U=u, c, \bar d, \bar s, \cdots, \\
  D=d, s, \bar u, \bar c, \cdots.
\end{align}
For the positively charged leptons, the d-type PDFs $g's$ change signs and the equations shown above remain the same.

We see that only 54 structure functions in total actually contribute at twist-3 level in the jet production SIDIS process. The other structure functions shown in Eqs. (\ref{f:Wuu})-(\ref{f:WLTT}) does not contribute because they are twist-4 effects.

\subsection{Azimuthal asymmetries from electron beam}

In addition to structure functions, we also calculate the azimuthal asymmetries results. In this part, we only show the results for the azimuthal asymmetries from the electron beam where only u-type contribute in this case. The results of the positron beam can be obtain in the similar way. We do not show them in this paper for simplicity.

We first present the definitions of the azimuthal asymmetries, e.g.,
\begin{align}
  \langle \sin\varphi \rangle_{U}=\frac{\int d\tilde\sigma \sin\varphi d\varphi}{\int d\tilde\sigma d\varphi},
\end{align}
for the unpolarized or longitudinally polarized target case, and
\begin{align}
  \langle \sin(\varphi-\varphi_S) \rangle_{T}=\frac{\int d\tilde\sigma \sin(\varphi-\varphi_S)d\varphi d\varphi_S}{\int d\tilde\sigma d\varphi d\varphi_S},
\end{align}
for the transversely polarized target case.
$d\tilde\sigma$ is used to denote $\frac{d\sigma}{dx dy d\psi d^2 k_\perp}$,
and $d\varphi_S\approx d\psi$
whose integration corresponds to take the average over the outgoing electron's azimuthal angle~\cite{Diehl:2005pc}.
The subscript $T$ denotes the polarization of the target. As before, we only present the electron beam case. The results of the positron case can be obtained similarly, we will not repeat it.
At the leading twist, there are six polarization dependent azimuthal asymmetries which are given by
\begin{align}
 & \langle \sin(\varphi-\varphi_S) \rangle_{T} = k_{\perp M} \frac{ f^\perp_{1T}}{2 f_1}, \\
 & \langle \cos(\varphi-\varphi_S) \rangle_{T} = - k_{\perp M} \frac{g^\perp_{1T}}{2 f_1}, \\
 & \langle \sin(\varphi-\varphi_{LT}) \rangle_{LT} = -k_{\perp M} \frac{g^\perp_{1LT}}{2f_1}, \\
 & \langle \cos(\varphi-\varphi_{LT}) \rangle_{LT} = - k_{\perp M} \frac{f^\perp_{1LT}}{2f_1}, \\
 & \langle \sin(2\varphi-\varphi_{TT}) \rangle_{TT} = - k_{\perp M}^2 \frac{g^\perp_{1TT}}{2f_1}, \\
 & \langle \cos(2\varphi-\varphi_{TT}) \rangle_{TT} = k_{\perp M}^2 \frac{f^\perp_{1TT}}{2f_1}.
\end{align}
All of the leading twist azimuthal asymmetries are generated by the correlations between the transverse polarization of the target and the transverse momentum of the parton inside the target. At twist-3, we have 18 azimuthal asymmetries. They are given by
\begin{align}
  & \langle \cos\varphi \rangle_{U} = -x\kappa_M k_{\perp M} \frac{T_{2}(y)}{T_{0}(y)} \frac{f^\perp}{f_1}, \\
  & \langle \sin\varphi \rangle_{U} = -x\kappa_M k_{\perp M} \frac{T_{2}(y)}{T_{0}(y)} \frac{g^\perp}{f_1}, \\
  & \langle \cos\varphi \rangle_{L} = -x\kappa_M k_{\perp M} \frac{T_{2}(y)}{T_{0}(y)} \frac{f^\perp -\lambda_h g^\perp_L}{f_1}, \\
  & \langle \sin\varphi \rangle_{L} = -x\kappa_M k_{\perp M} \frac{T_{2}(y)}{T_{0}(y)} \frac{g^\perp + \lambda_h f^\perp_L}{f_1}, \\
  & \langle \cos\varphi \rangle_{LL} = -x\kappa_M k_{\perp M} \frac{T_{2}(y)}{T_{0}(y)} \frac{f^\perp + S_{LL}f^\perp_{LL}}{f_1}, \\
  & \langle \sin\varphi \rangle_{LL} = -x\kappa_M k_{\perp M} \frac{T_{2}(y)}{T_{0}(y)} \frac{g^\perp + S_{LL}g^\perp_{LL}}{f_1}, \\
  & \langle \cos\varphi_S \rangle_{T} =  x\kappa_M \frac{T_{2}(y)}{T_{0}(y)} \frac{g_T}{f_1}, \\
  & \langle \sin\varphi_S \rangle_{T} = -x\kappa_M \frac{T_{2}(y)}{T_{0}(y)} \frac{f_T}{f_1}, \\
  & \langle \cos(2\varphi-\varphi_S) \rangle_{T} = x\kappa_M k_{\perp M}^2 \frac{T_{2}(y)}{T_{0}(y)} \frac{g^\perp_T}{f_1}, \\
  & \langle \sin(2\varphi-\varphi_S) \rangle_{T} = -x\kappa_M k_{\perp M}^2 \frac{T_{2}(y)}{T_{0}(y)} \frac{f^\perp_T}{f_1}, \\
  & \langle \cos\varphi_{LT} \rangle_{LT} = -x\kappa_M \frac{T_{2}(y)}{T_{0}(y)} \frac{f_{LT}}{f_1}, \\
  & \langle \sin\varphi_{LT} \rangle_{LT} = -x\kappa_M \frac{T_{2}(y)}{T_{0}(y)} \frac{g_{LT}}{f_1}, \\
  & \langle \cos(2\varphi-\varphi_{LT}) \rangle_{U,LT} = -x\kappa_M k_{\perp M}^2 \frac{T_{2}(y)}{T_{0}(y)} \frac{f^\perp_{LT}}{2f_1}, \\
  & \langle \sin(2\varphi-\varphi_{LT}) \rangle_{LT} = -x\kappa_M k_{\perp M}^2 \frac{T_{2}(y)}{T_{0}(y)}
  \frac{g^\perp_{LT}}{2f_1}, \\
  & \langle \cos(\varphi-2\varphi_{TT}) \rangle_{TT} =  x\kappa_M k_{\perp M} \frac{T_{2}(y)}{T_{0}(y)} \frac{f_{TT}}{f_1}, \\
  & \langle \sin(\varphi-2\varphi_{TT}) \rangle_{TT} = -x\kappa_M k_{\perp M} \frac{T_{2}(y)}{T_{0}(y)} \frac{g_{TT}}{f_1}, \\
  & \langle \cos(3\varphi-3\varphi_{TT}) \rangle_{TT} = x\kappa_M k_{\perp M}^3 \frac{T_{2}(y)}{T_{0}(y)} \frac{f^\perp_{TT}}{2f_1}, \\
  & \langle \sin(3\varphi-2\varphi_{TT}) \rangle_{TT} = x\kappa_M k_{\perp M}^3 \frac{T_{2}(y)}{T_{0}(y)} \frac{g^\perp_{TT}}{2f_1}.
\end{align}
These azimuthal asymmetries can be measured in the jet production charged current SIDIS to extract the corresponding twist-3 PDFs.

\subsection{Charge asymmetries}

Charge asymmetry is assumed to be valid in parton model for a few reasons, see ref.~\cite{Londergan:2009kj}. It provides a powerful tool to study and understand strong interaction systems \cite{Miller:1990iz,Miller:2006tv}. For example, model calculation can help to understand the mass difference between the $u$ and $d$ quarks \cite{Hutauruk:2018zfk}. In this subsection we consider the charge asymmetries induced by the exchange of the incoming leptons ($e^- \leftrightarrow e^+$) in the lepton scattering process.

It is convenient to consider the inclusive DIS which can be obtained by integrating over $d^2k_\perp$ in Eqs. (\ref{f:crosst2})-(\ref{f:crosst3}). To calculate the charge asymmetries, we first introduce the $plus$ and $minus$ cross sections which are given by
\begin{align}
  d\sigma^P_{in}=\frac{d\sigma_{in}(e^-)}{dx dy d\psi} + \frac{d\sigma_{in}(e^+)}{dx dy d\psi}, \label{f:crossplus}\\
  d\sigma^M_{in}=\frac{d\sigma_{in}(e^-)}{dx dy d\psi} - \frac{d\sigma_{in}(e^+)}{dx dy d\psi}, \label{f:crossminus}
\end{align}
where superscripts $P, M$ denote the plus and minus cross sections calculated by  Eqs. (\ref{f:crossplus}), (\ref{f:crossminus}), respectively. The explicit expressions of the plus and minus cross sections are

\begin{align}
  d\sigma_{in}^P =  \frac{\alpha_{\rm{em}}^2 }{y Q^2}8\chi_W\biggl\{ T_0(y)& \left[\left(f^U_1(x)+f^D_1(x)\right) +S_{LL}\left(f^U_{1LL}(x)+f^D_{1LL}(x)\right)-\lambda_h \left(g^U_{1L}(x)-g^D_{1L}(x)\right)\right] \nonumber\\
  -2x\kappa_M T_2(y)\Big\{|S_T| &\left[\sin\varphi_S \left(f^U_T(x)+f^D_T(x)\right)- \cos\varphi_S \left(g^U_T(x)-g^D_T(x)\right) \right]  \nonumber\\
  +|S_{LT}|& \left[\sin\varphi_{LT} \left(g^U_{LT}(x)-g^D_{LT}(x) \right) +\cos\varphi_{LT} \left( f^U_{LT}(x)+f^D_{LT}(x)\right)\right]\Big\}  \biggr\}. \label{f:crossplusin} \\
  d\sigma_{in}^M =  \frac{\alpha_{\rm{em}}^2 }{y Q^2}8\chi_W\biggl\{ T_0(y)& \left[\left(f^U_1(x)-f^D_1(x)\right) +S_{LL}\left(f^U_{1LL}(x)-f^D_{1LL}(x)\right)-\lambda_h \left(g^U_{1L}(x)+g^D_{1L}(x)\right)\right] \nonumber\\
  -2x\kappa_M T_2(y)\Big\{|S_T| &\left[\sin\varphi_S \left(f^U_T(x)-f^D_T(x)\right)- \cos\varphi_S \left(g^U_T(x)+g^D_T(x)\right) \right]  \nonumber\\
  +|S_{LT}|& \left[\sin\varphi_{LT} \left(g^U_{LT}(x)+g^D_{LT}(x) \right) +\cos\varphi_{LT} \left( f^U_{LT}(x)-f^D_{LT}(x)\right)\right]\Big\}  \biggr\}. \label{f:crossminusin}
\end{align}

Here we introduce the definitions of the charge asymmetries. For the unpolarized target, we have
\begin{align}
  A_\sigma^{CA}=\frac{d\sigma^{P/M}_{in,\sigma}(\sigma=0)}{d\sigma_U^{\gamma\gamma}}, \label{f:ChargeAUnDef}
\end{align}
where the subscript $\sigma$ denotes the target polarization, superscript $CA$ denotes charge asymmetry, $d\sigma_U^{\gamma\gamma}$ denotes the unpolarized electromagnetic cross section. For the polarized target, the definition is defined as
\begin{align}
  A_\sigma^{CA}=\frac{d\sigma^{P/M}_{in,\sigma}(+\sigma)-d\sigma^{P/M}_{in,\sigma}(-\sigma)}{d\sigma_U^{\gamma\gamma}}. \label{f:ChargeADef}
\end{align}

First of all, we take the target as unpolarized. According to the definition, we have
\begin{align}
 & A_U^{CA,P}=\frac{8\chi_W T_0(y)\left(f^U_1(x)+f^D_1(x)\right) }{e_q^2 A(y) f^q_1(x)}, \label{f:AUP} \\
 & A_U^{CA,M}=\frac{8\chi_W T_0(y)\left(f^U_1(x)-f^D_1(x)\right) }{e_q^2 A(y) f^q_1(x)}.
\end{align}
where the index mark $q$ denotes the flavor of quark. We note that a summation over the corresponding flavor in both the numerator and denominator is understood.

For the longitudinal polarized target, we have
\begin{align}
 & A_L^{CA,P}=-\frac{8\chi_W T_0(y)\left(g^U_{1L}(x)-g^D_{1L}(x)\right) }{e_q^2 A(y) f^q_1(x)}, \\
 & A_L^{CA,M}=-\frac{8\chi_W T_0(y)\left(g^U_{1L}(x)+g^D_{1L}(x)\right) }{e_q^2 A(y) f^q_1(x)}, \\
 & A_{LL}^{CA,P}=\frac{8\chi_W T_0(y)\left(f^U_{1LL}(x)+f^D_{1LL}(x)\right) }{e_q^2 A(y) f^q_1(x)}, \\
 & A_{LL}^{CA,M}=\frac{8\chi_W T_0(y)\left(f^U_{1LL}(x)-f^D_{1LL}(x)\right) }{e_q^2 A(y) f^q_1(x)}.
\end{align}

For the transverse polarized target, we have
\begin{align}
  & A_{T,x}^{CA,P}=\frac{16x\kappa_M \chi_W T_2(y)\left(g^U_{T}(x)-g^D_{T}(x)\right)}{e_q^2 A(y) f_1(x)}, \\
  & A_{T,x}^{CA,M}=\frac{16x\kappa_M \chi_W T_2(y)\left(g^U_{T}(x)+g^D_{T}(x)\right)}{e_q^2 A(y) f_1(x)}, \\
  & A_{T,y}^{CA,P}=-\frac{16x\kappa_M \chi_W T_2(y)\left(f^U_{T}(x)+f^D_{T}(x)\right)}{e_q^2 A(y) f_1(x)}, \\
  & A_{T,y}^{CA,M}=-\frac{16x\kappa_M \chi_W T_2(y)\left(f^U_{T}(x)-f^D_{T}(x)\right)}{e_q^2 A(y) f_1(x)}, \\
  & A_{LT,x}^{CA,P}=-\frac{16x\kappa_M \chi_W T_2(y)\left(f^U_{LT}(x)+f^D_{LT}(x)\right)}{e_q^2 A(y) f_1(x)}, \\
  & A_{LT,x}^{CA,M}=-\frac{16x\kappa_M \chi_W T_2(y)\left(f^U_{LT}(x)-f^D_{LT}(x)\right)}{e_q^2 A(y) f_1(x)}, \\
  & A_{LT,y}^{CA,P}=-\frac{16x\kappa_M \chi_W T_2(y)\left(g^U_{LT}(x)-g^D_{LT}(x)\right)}{e_q^2 A(y) f_1(x)}, \\
  & A_{LT,y}^{CA,M}=-\frac{16x\kappa_M \chi_W T_2(y)\left(g^U_{LT}(x)+g^D_{LT}(x)\right)}{e_q^2 A(y) f_1(x)}. \label{f:ALTM}
\end{align}

From Eqs. (\ref{f:AUP})-(\ref{f:ALTM}) we can see that asymmetries are divided into two parts by the $plus$ and $minus$ cross sections. According to the definitions, the $plus$ asymmetries can be used to determine the plus combination of the unpolarized PDFs ($f's$) and/or the minus combination of the longitudinal polarized PDFs ($g's$). On the contrary, the $minus$ asymmetries can be used to determine the minus combination of the unpolarized PDFs ($f's$) and/or the $plus$ combination of the longitudinal polarized PDFs ($g's$). So these charge asymmetries defined here are important and convenient to determine the corresponding PDFs.

Charge asymmetries given in this part combine the electro-weak and QCD theories. Measuring these asymmetries are important ways to examine electroweak and QCD theories simultaneously.

\section{Summary} \label{sec:summary}

In this paper, we present a complete and systematic calculation of the current jet production SIDIS process at the EIC.
Only the charged current interaction is considered in the calculation. We first presented the general form of the differential cross section of this process in terms of structure functions by making full kinematical analysis.
In the parton model the calculations are carried out by applying the collinear expansion where the multiple gluon scattering is taken into account and gauge links are obtained systematically and automatically.
The calculations are limited in the leading order twist-3 level.
By matching the differential cross sections given by structure functions and gauge invariant PDFs, we obtained 54 structure functions which contribute in the jet production SIDIS process. The twist-4 structure functions were not considered.
We also presented the azimuthal asymmetries results. There are 6 leading twist azimuthal asymmetries and 18 twist-3 azimuthal asymmetries in total. All of them are presented in terms of gauge invariant PDFs.
By introducing the $plus$ and $minus$ cross sections, we calculated the charge asymmetries for both the unpolarized and polarized targets, respectively.

\section*{Acknowledgements}
The author thanks Prof. Zuo-Tang Ling and Kai-Bao Chen very much for his helpful suggestions. This work was supported the National Laboratory Foundation (Grant No. 6142004180203).

\end{document}